\documentclass{article}
\pdfoutput=1 

\usepackage[final]{neurips_2019}

\usepackage[utf8]{inputenc} 
\usepackage[T1]{fontenc}    
\usepackage{hyperref}       
\usepackage{url}            
\usepackage{booktabs}       
\usepackage{amsmath}    	
\usepackage{amsfonts}       
\usepackage{amssymb}    	
\usepackage{nicefrac}       
\usepackage{microtype}      
\usepackage{bm}
\usepackage{graphicx}
\usepackage{subcaption}

\DeclareMathOperator*{\argmax}{arg\,max}
\DeclareMathOperator*{\argmin}{arg\,min}
\DeclareMathOperator*{\prox}{prox}
\newcommand{\scarlet}{{\sc scarlet}}

\title{Hybrid Physical-Deep Learning Model for Astronomical Inverse Problems}

\author{%
  Fran\c{c}ois Lanusse \\
  Berkeley Center for Cosmological Physics\\
  Berkeley Institute for Data Science \\
  University of California, Berkeley\\
  Berkeley, CA 94709 \\
  \texttt{flanusse@berkeley.edu} \\
  \And
  Peter Melchior \\
  Department of Astrophysical Sciences\\
  Center for Statistics and Machine Learning\\
  Princeton University\\
  Princeton, NJ 08544, USA\\
  \texttt{peter.melchior@princeton.edu}\\
  \And
  Fred Moolekamp \\
  LSST Project Management Office \\
  Tucson, AZ, USA \\
  \medskip
  Department of Astrophysical Sciences\\
  Princeton University\\
  Princeton, NJ 08544, USA\\
  \texttt{fredem@astro.princeton.edu}
}

\begin{document}

\maketitle

\begin{abstract}
  We present a Bayesian machine learning architecture that combines a physically motivated parametrization and an analytic error model for the likelihood with a deep generative model providing a powerful data-driven prior for complex signals.
  This combination yields an interpretable and differentiable generative model, allows the incorporation of prior knowledge, and can be utilized for observations with different data quality without having to retrain the deep network.
  We demonstrate our approach with an example of astronomical source separation in current imaging data, yielding a physical and interpretable model of astronomical scenes.
\end{abstract}

\section{Introduction}

Deep Learning is extremely efficient at solving a wide range of inverse problems, but what is gained in performance is often lost in interpretability due the black box nature of deep neural networks. In scientific applications however, the interpretability of the solution is often paramount, as is the ability to imbue pre-existing physical knowledge directly into the model. To this end, we propose an hybrid model based on solving for the Maximum A Posteriori (MAP) solution of an inverse problem, where a physical model is used to describe the forward data acquisition process, and a deep generative model with explicit likelihood is used to provide a complex data-driven signal prior. 

\paragraph{Related Work} With the success of deep learning in many classical imaging problems \citep[e.g.][]{Dong2016}, a significant amount of effort has been aimed at linking deep learning successes back to the classical inverse problems literature. Several avenues have been explored, e.g. learning a denoiser as an implicit proximal operator \citep{Meinhardt2017}; using a generative model as deep image prior \citep{Lempitsky2018}; learning a convolutional dictionary and proximal operator \citep{Yang2016}; or learning implicitly both the prior and the inference algorithm itself \citep{Putzky2017}. Finally, most closely related to our work, \citep{Dave2018} introduces the idea of modeling the prior explicitly at the pixel-level using a deep generative model.

\section{Problem statement}
\label{sec:problem_statement}

We consider a general linear inverse problems of the form:
\begin{equation}
\label{eq:lip}
    \bm{y} = \mathbf{A} \bm{x} + \bm{n} \;,
\end{equation}
where $\bm{y}$ are the observations, $\bm{x}$ is the unknown signal to recover, $\mathbf{A}$ is a linear degradation operator, $\bm{n}$ is some observational noise. This problem describes a wide range of applications from MRI to radio-interferometry through different choices of the operator $\mathbf{A}$.

In the Bayesian approach to inverse problems, the posterior $p(\bm{x} | \bm{y})$ can be expressed as
\begin{equation}
    p(\bm{x} | \bm{y}) \propto p(\bm{y} | \bm{x}) \  p(\bm{x}).
\end{equation}%
\begin{itemize}
    \item The \textit{data likelihood} term $p(\bm{y} | \bm{x})$ encodes our physical understanding of the forward process that leads to the observations. We assume here that for a given $\bm{x}$, this term can be evaluated explicitly given our physical model.
    \item The \textit{prior} term $p(\bm{x})$ encodes our prior knowledge on the solution we seek to recover. This prior can be informed by prior experiments, complementary data, or physical considerations. A tractable expression of this prior can often  be obtained only for simple signal classes.
\end{itemize}

While the Bayesian solution to such inverse problem is the full posterior $p(\bm{x} | \bm{y})$, in many practical applications the full distribution is typically reduced to a single point estimate, i.e. the Maximum A Posteriori (MAP) solution
\begin{equation}
    \bm{x}_{MAP} = \argmax\limits_{x} \ \log p(\bm{y} | \bm{x})  + \log p( \bm{x} ) \;.
    \label{eq:MAP}
\end{equation}

We will exploit this separable representation, limiting the use of deep learning to the prior term $p(\bm{x})$. 

\paragraph{Model for the data likelihood} With a given operator $\mathbf{A}$, the likelihood term is completely characterized by the noise model for $\bm{n}$. We assume a Gaussian noise model, i.e. $\bm{n} \sim \mathcal{N} (0, \mathbf{\Sigma})$ where $\mathbf{\Sigma}$ is the noise covariance. In this case $\log p(\bm{y} | \bm{x}) = - \frac{1}{2}  \lVert \bm{y} - \mathbf{A} \bm{x} \rVert_\mathbf{\Sigma^{-1}}^2 + cst$, where $\lVert \bm{x} \rVert_\mathbf{\bf{M}}^2 = \bm{x}^\top \mathbf{M} \bm{x}$.

\paragraph{Deep generative models as complex data priors} Models like Variational AutoEncoders (VAEs) \citep{Kingma2013} and Generative Adversarial Networks (GANs) \citep{Goodfellow2014} have been extremely successful, but they do not provide an explicit likelihood $p(\bm{x})$. Instead we choose to rely on pixel autoregressive models \citep{Oord2016, Salimans2017, Chen2018} which provide an explicit likelihood, factorized into separate conditional distributions $p_\theta(\bm{x}) = \Pi_{i} p_\theta(x_i | x_0 \ldots x_{i-1})$ where $\theta$ are the weights of the model. These models achieve state of the art performance, are stable during training, and do not suffer from mode collapse (contrary to GANs). Our prior model is trained on uncorrupted examples of data $\bm{x}$, which may come from simulations or from high-fidelity observations.

\paragraph{Physical constraints as proximal regularization terms} In this approach, most of the prior information stems from the data-driven deep generative model $p_\theta(\bm{x})$. However, additional physical constraints on the solution can applied by adding regularization terms to the prior: $\log p(\bm{x}) = \log p_\theta(\bm{x}) + \sum_j R_j(\bm{x})$. The regularizers may be non-differentiable as long as they can be expressed by their proximal operators $\prox_{R_j}(x)$.  As an example, the flux of astronomical sources is a positive quantity, we may therefore want to impose a non-negativity constraint $\iota_{>0}(\bm{x})$ on the solution, using the associated proximal operator $\prox_{\iota_{>0}} (x) = \max(0, x)$.

Combining all these elements, we can characterize our $x_{MAP}$ solution as the minimum of the following loss function: 
\begin{equation}
    \mathcal{L} = \frac{1}{2}  \lVert \bm{y} - \mathbf{A} \bm{x} \rVert_\mathbf{\Sigma^{-1}}^2 - \log p_\theta(\bm{x}) + \sum_i R_i(\bm{x}) \equiv f(\bm{x}) + g(\bm{x}) + r(\bm{x}) \;.
    \label{eq:loss_function}
\end{equation}
The two first components of this loss are differentiable and therefore amenable to gradient descent.
Due to the presence of non-differentiable regularizers, the optimization makes use of the iterative Proximal Gradient Method (also known as forward-backward splitting \citep{Combettes2005})
\begin{equation}
\label{eq:pgm}
    \bm{x}_{t+1} = \prox_{\lambda_t\, r}\left(\bm{x}_t - \lambda_t\nabla (f + g)(\bm{x}_t)\right),
\end{equation}
which converges to a minimum of $\mathcal{L}$ if the step size $\lambda$ is smaller than $2/L$, where L is the Lipschitz constant of the gradient term.

\section{Application: deblending galaxy images}

\begin{figure}
    \centering
    \begin{subfigure}[b]{0.3\textwidth}
        \includegraphics[width=\linewidth]{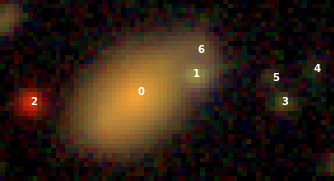}
        \caption{Observed Scene}
    \end{subfigure}
    \begin{subfigure}[b]{0.3\textwidth}
    \includegraphics[width=\linewidth]{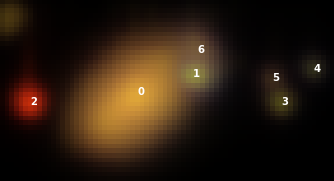}
        \caption{Sky Model}
        \label{fig:sky_model}
    \end{subfigure}
    \begin{subfigure}[b]{0.3\textwidth}
    \includegraphics[width=\linewidth]{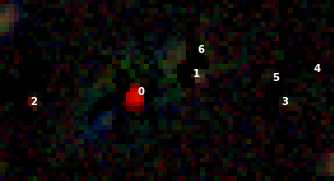}
        \caption{Residuals}
    \end{subfigure}\\
    \begin{subfigure}[b]{\textwidth}
        \centering
        \includegraphics[width=0.3\linewidth]{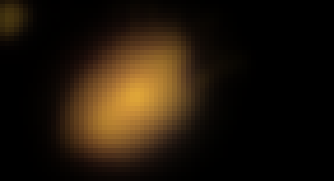}
        \includegraphics[width=0.3\linewidth]{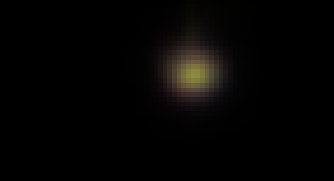}
        \includegraphics[width=0.3\linewidth]{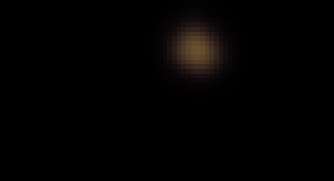}
        \caption{Individual components of the model}
        \label{fig:sky_components}
    \end{subfigure}
    \caption{\small Deblending of a scene from the HSC imaging survey using the proposed model. The sky model (b), composed of individual sources (marked in white) with their own morphology and SED, is fitted to the observations (a). A subset of the individual recovered components are shown in (d).}
    \label{fig:hsc_example}\vspace{-0.5cm}
\end{figure}

Modern wide-field cosmological surveys cover large areas of the sky with ever increasing imaging quality \citep[e.g.][]{LSST2009,Laureijs2011}.
They often seek to simultaneously address many scientific goals, e.g. mapping the large scale structure of the Universe to answer fundamental questions on the nature of Dark Matter and Dark Energy. One outstanding challenge faced by all modern imaging surveys is the overlap of several sources (stars and galaxies) on the sky, so-called "blending". It complicates the measurement of properties of individual members of the blend. This problem constitutes a typical instance of blind source separation, which attempts to separate all components of a blended scene without \emph{a priori} knowledge of their nature (see \autoref{fig:hsc_example}). A GAN-based approach to deblending was introduced in \citep{Reiman2019} but suffers from the typical limitations of black-box deep learning models, i.e. it cannot account for different observing conditions or noise levels without retraining the full model, is limited to separating two components, and has no test-time flux preservation.

\begin{figure}
    \centering
    \begin{subfigure}[b]{0.3\textwidth}
        \includegraphics[width=\linewidth]{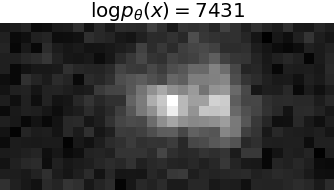}
        \caption{Simulated blend}
        \label{fig:sim_blend}
    \end{subfigure}
    \begin{subfigure}[b]{0.3\textwidth}
    \includegraphics[width=\linewidth]{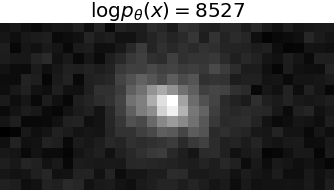}
        \caption{True component 1}
        \label{fig:cosmos_1}
    \end{subfigure}
    \begin{subfigure}[b]{0.3\textwidth}
    \includegraphics[width=\linewidth]{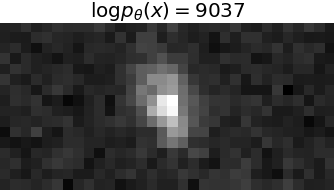}
        \caption{True component 2}
        \label{fig:cosmos_2}
    \end{subfigure}\\
    \begin{subfigure}[b]{0.3\textwidth}
        \includegraphics[width=\linewidth]{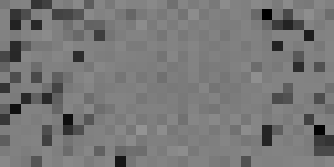}
        \caption{residuals}
    \end{subfigure}
    \begin{subfigure}[b]{0.3\textwidth}
        \includegraphics[width=\linewidth]{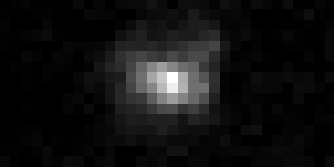}
        \caption{Recovered component 1}
        \label{fig:rec_1}
    \end{subfigure}
    \begin{subfigure}[b]{0.3\textwidth}
        \includegraphics[width=\linewidth]{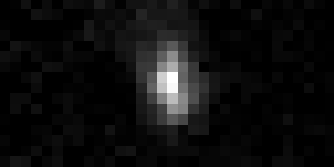}
        \caption{Recovered component 2}
        \label{fig:rec_2}
    \end{subfigure}
    \caption{\small Artificial blend experiment, simulated from HST observations, demonstrating the ability of the prior to disentangle
    sources from their morphologies alone. The log likelihood evaluated on the trained prior for the blended and isolated sources is reported in the top row: a blended object (a) has lower log likelihood than isolated objects (b, c) under the prior. This allows us to separate (e) and (f) when given (a).}
    \label{fig:hst_example} \vspace{-0.5cm}
\end{figure}

We propose to address the deblending problem using the analytic model of the scene introduced by \cite{Melchior2018}. Given an astronomical scene $\bm{y} \in \mathbb{R}^{B\times N\times N}$ observed in $B$ bands (i.e. using multiple filters), each source $k$ in the scene is modeled with a non-parametric shape $S_k \in \mathbb{R}^{N \times N}$ and an amplitude $A_k \in \mathbb{R}^{B}$, the so-called Spectral Energy Distribution (SED), which determines how bright the object will appear in each band.
Multiple sources contribute additively to the scene, which is correct in the absence of absorbers, e.g. inter-stellar dust. 
The forward model also needs to account for degradation of the image caused by the atmosphere and the instrumental optics. This can be described as a band-wise 2D convolution by a Point Spread Function (PSF). We denote $\mathbf{P}$ as block-diagonal linear operator implementing the convolution in each band by the appropriate PSF. It acts as the operator $\mathbf{A}$ in \autoref{eq:lip}. Our full physical model for the scene can now be expressed as:
\begin{equation}
    \bm{y} = \mathbf{P}\sum_{k=1}^K A_k^T \times S_k + \bm{n}
    \label{eq:forward_model}
\end{equation}
where $\bm{n} \in \mathbb{R}^{B\times N\times N}$ is typically assumed to be Gaussian noise with covariance $\Sigma$. The deblending problem is to recover an estimate of both the morphology $S_k$ and the SED $A_k$ of each component of the blend, subject to additional constraints such as positivity of the source emission ($S_k > 0$ and $A_k > 0$). Applying the framework described in \autoref{sec:problem_statement}, we solve the optimization problem
\begin{equation}
    \argmin_{S_k, A_k} \frac{1}{2}  \lVert \bm{y} - \mathbf{P}\sum_{k=1}^K A_k^T \times S_k  \rVert_{\mathbf{\Sigma^{-1}}}^2 + \sum_k^K \log p_\theta( S_k ) +  \iota_{>0}(A_k) +  \iota_{>0}(S_k),
    \label{eq:optim_deblending}
\end{equation}
by a block-wise application of \autoref{eq:pgm} to every optimization variable.
We base the morphology prior $p_\theta$ on the PixelCNN++ model \citep{Salimans2017}, which we adjust for continuous signals by using a simple Gaussian model for the conditional distribution $p_\theta(x_i | x_0 \ldots x_{i-1})$.
The prior $p_\theta$ is trained on an existing set of high-resolution images of isolated galaxies \citep{mandelbaum, Mandelbaum2014, Rowe2015} from the Hubble Space Telescope (HST)/Advanced Camera for Surveys (ACS) COSMOS survey \citep{Koekemoer2007}. These single-band high-resolution images from HST are reconvolved with a uniform reference PSF and resampled to match the pixel scale of the test survey, the Hyper Suprime-Cam (HSC) survey from the Subaru 8.2 meter telescope on Maunakea, Hawaii \citep{Aihara2018}. \autoref{fig:cosmos_1} and \autoref{fig:cosmos_2} show examples of two isolated galaxies obtained by this procedure, annotated with the log likelihood $\log p_\theta(x)$ of the trained prior.  \autoref{fig:sim_blend} shows a simulated blend obtained by adding these two isolated galaxies. The log likelihood under the prior for both galaxies combined is lower than for each isolated galaxy, demonstrating that the morphology prior provides information that can be leveraged for deblending.

\begin{figure}[h]
    \centering
    \begin{subfigure}[b]{0.3\textwidth}
        \includegraphics[width=\linewidth]{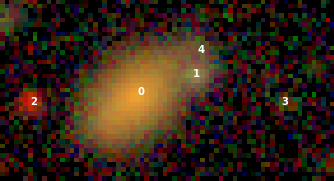}
        \caption{Simulated noisy scene}
    \end{subfigure}
    \begin{subfigure}[b]{0.3\textwidth}
    \includegraphics[width=\linewidth]{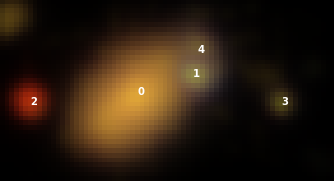}
        \caption{Sky model}
    \end{subfigure}
    \begin{subfigure}[b]{0.3\textwidth}
    \includegraphics[width=\linewidth]{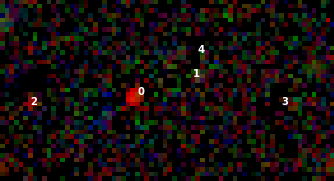}
        \caption{Residuals}
    \end{subfigure}\\
    \begin{subfigure}[b]{\textwidth}
        \centering
        \includegraphics[width=0.3\linewidth]{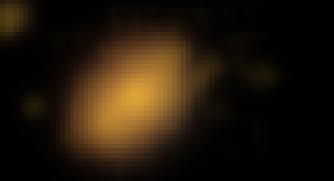}
        \includegraphics[width=0.3\linewidth]{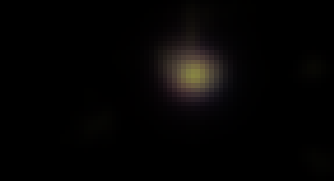}
        \includegraphics[width=0.3\linewidth]{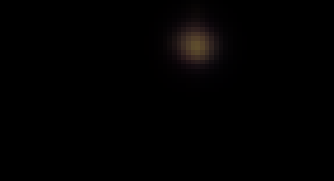}
        \caption{Individual components of the model}
    \end{subfigure}
    \caption{\small Same as \autoref{fig:hsc_example}, but with artificially increased noise (RMS $\times\ 3$).}
    \label{fig:hsc_example_noise}
\end{figure}

Equipped with the prior on single-galaxy morphology, we tackle the blended scene in \autoref{fig:hsc_example} from the first public HSC data. The observations, made in $B=5$ different filters, are modelled with \autoref{eq:forward_model}, where the number of sources $K$ and a first guess of their positions is provided by an external detection algorithm. We infer the parameters of the scene by solving \autoref{eq:optim_deblending}, yielding a sky model (\autoref{fig:sky_model}) which can be separated into its individual components (\autoref{fig:sky_components}). The model creates an excellent fit to morphologies and SEDs despite the strong overlap of several sources. The residuals are dominated by an undetected nuclear component in the brightest galaxy, which could be modeled by adding another component there. Finally, our model analytically accounts for different observing conditions. We provide in \autoref{sec:appendix_scarlet} a comparison to the state-of-the-art \scarlet\ deblender \citep{Melchior2018} on this scene to highlight the benefits of the deep morphology prior. In \autoref{fig:hsc_example_noise} we show the same blended scene, but we artificially increased the noise RMS by a factor of 3. By adjusting the noise covariance in \autoref{eq:optim_deblending}, the methods recovers a very similar result without the need to retrain the deep learning prior. We could also address e.g. additional blurring from a wider PSF or resampling to lower resolution with the same approach. \vspace{-0.25cm}

\section{Conclusion} \vspace{-0.25cm}

We have presented a hybrid Bayesian framework for inverse problems that combines analytic forward modeling for the likelihood with deep generative models for complex data-driven signal priors. This approach makes explicit use of physically motivated problem structure and prior knowledge from high-quality observations. When applied to the blind-source separation problem of galaxy blending, we can retrieve multi-components models of astronomical scenes that are by construction robust to changes in observational conditions.

\bibliographystyle{apalike}
\bibliography{biblio}

\appendix

\section{Comparison to the \scarlet\ deblender}
\label{sec:appendix_scarlet}

In this appendix, we compare the proposed method with the baseline deblending algorithm \scarlet, which constitutes a state-of-the-art deblender for ground-based images \citep{Melchior2018}. \scarlet\ uses the same parameterization and loss function; in fact the work described here uses the same code base and only differs in the assumptions about galaxy morphologies.

In its default configuration, \scarlet\ assumes every $S_k$ to be non-negative, symmetric under rotation of 180$^\circ$ and monotonically decreasing away from the center.
These hard constraints can directly be enforced through proximal mappings in \autoref{eq:loss_function} and have been found successful as regularizers of the deblending problem for ground-based images.
They do not perform well on complex and irregular galaxies, which is the original motivation for the present work: replacing analytic, heuristic constraints by a data-driven deep morphology prior. 

\begin{figure}
    \centering
    \begin{subfigure}[b]{0.3\textwidth}
        \includegraphics[width=\linewidth]{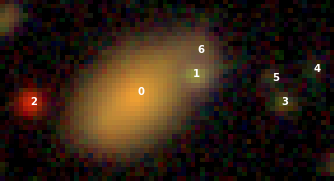}
        \caption{Simulated noisy scene}
    \end{subfigure}
    \begin{subfigure}[b]{0.3\textwidth}
    \includegraphics[width=\linewidth]{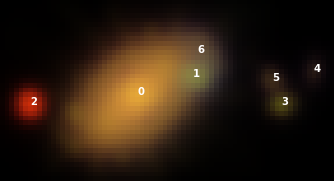}
        \caption{\scarlet\ sky model}
    \end{subfigure}
    \begin{subfigure}[b]{0.3\textwidth}
    \includegraphics[width=\linewidth]{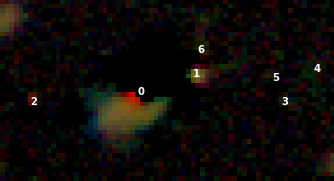}
        \caption{Residuals}
    \end{subfigure}\\
    \begin{subfigure}[b]{\textwidth}
        \centering
        \includegraphics[width=0.3\linewidth]{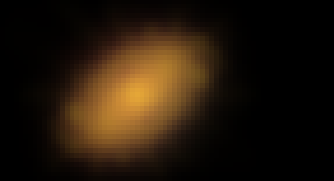}
        \includegraphics[width=0.3\linewidth]{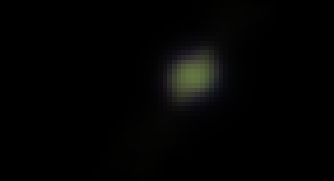}
        \includegraphics[width=0.3\linewidth]{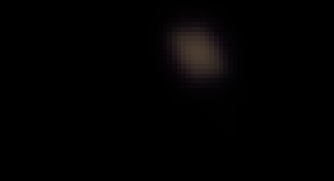}
        \caption{Individual components of the model recovered by \scarlet}
    \end{subfigure}\\
    \begin{subfigure}[b]{\textwidth}
        \centering
        \includegraphics[width=0.3\linewidth]{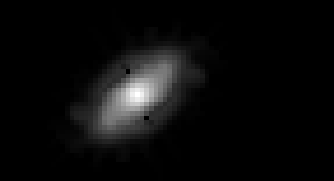}
        \includegraphics[width=0.3\linewidth]{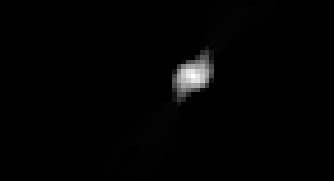}
        \includegraphics[width=0.3\linewidth]{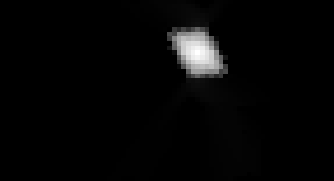}
        \caption{Individual morphology components recovered with \scarlet}
    \end{subfigure}\\
    \begin{subfigure}[b]{\textwidth}
        \centering
        \includegraphics[width=0.3\linewidth]{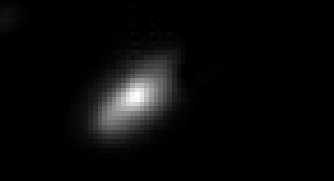}
        \includegraphics[width=0.3\linewidth]{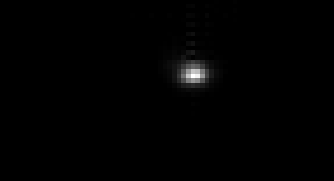}
        \includegraphics[width=0.3\linewidth]{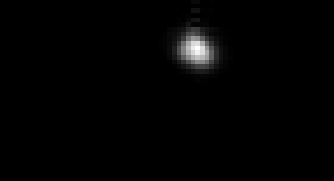}
        \caption{Individual morphology components recovered by our proposed method}
    \end{subfigure}
    \caption{\small Similar to \autoref{fig:hsc_example}, but using the deblender \scarlet\ with default settings. We also compare in (e) and (f) the deconvolved morphology components ($S_k$ in \autoref{eq:forward_model}) recovered using the strict monotonicity and symmetry constraints in \scarlet\ with the deep morphology priors from this work.}
    \label{fig:hsc_example_scarlet}
\end{figure}

\autoref{fig:hsc_example_scarlet} show the result of running baseline \scarlet\ on the same data set.
The residuals are significantly larger than in \autoref{fig:hst_example} due to the inadequacies of the strict symmetry and monotonicity assumptions.
This can also be seen by directly comparing the recovered deconvolved morphologies (lower panels of \autoref{fig:hsc_example_scarlet}).
The symmetry constraint can lead to artifacts in the direction of a nearby source, in this example the model of source 0 is influenced by source 1.
The morphologies recovered under the deep morphology prior are by construction realistic and do not exhibit such obvious artifacts.

While this comparison remains qualitative, it illustrates that the deep morphology prior addresses one main limitation of the \scarlet\ algorithm. A thorough study of our extension for the science cases of the upcoming LSST survey \citep{LSST2009} will be the main focus of an upcoming science paper geared towards the astronomical community.

\end{document}